\documentclass[journal=nalefd,manuscript=letter,layout=onecolumn,%
doi=true,maxauthors=0]{achemso}
\setkeys{acs}{etalmode=truncate,maxauthors=0}
\usepackage[version=3]{mhchem} 

\author{E. V. Calman}
\altaffiliation{These authors contributed equally}
\author{L. H. Fowler-Gerace}
\email{lefowler@physics.ucsd.edu}
\altaffiliation{These authors contributed equally}
\author{D. J. Choksy}
\author{L. V. Butov}
\phone{+1 (858) 525-1944}
\affiliation[University of California, San Diego
]{Department of Physics, University of California at San Diego, La Jolla, CA 92093, USA}
\author{D. E. Nikonov}
\author{I. A. Young}
\affiliation[Intel Corporation]
{Components Research, Intel Corporation, Hillsboro, OR 97124 USA}
\author{S. Hu}
\author{A. Mishchenko}
\author{A. K. Geim}
\affiliation[University of Manchester]
{School of Physics and Astronomy, University of Manchester, Oxford Road, Manchester M13 9PL, UK}

\title[Indirect excitons and trions]
  {Indirect excitons and trions in MoSe$_2$/WSe$_2$ van der Waals heterostructures}

\abbreviations{}

\begin{document}
\begin{abstract}
Indirect excitons (IX) in semiconductor heterostructures are bosons, which can cool below the temperature of quantum degeneracy and can be effectively controlled by voltage and light. IX quantum Bose gases and IX devices were explored in GaAs heterostructures where an IX range of existence is limited to low temperatures due to low IX binding energies. IXs in van der Waals transition-metal dichalcogenide (TMD) heterostructures are characterized by large binding energies giving the opportunity for exploring excitonic quantum gases and for creating excitonic devices at high temperatures. TMD heterostructures also offer a new platform for studying single-exciton phenomena and few-particle complexes. In this work, we present studies of IXs in MoSe$_2$/WSe$_2$ heterostructures and report on two IX luminescence lines whose energy splitting and temperature dependence identify them as neutral and charged IXs. The experimentally found binding energy of the indirect charged excitons, i.e. indirect trions, is close to the calculated binding energy of 28 meV for negative indirect trions in TMD heterostructures [Deilmann, Thygesen, Nano Lett. 18, 1460 (2018)]. We also report on the realization of IXs with a luminescence linewidth reaching 4~meV at low temperatures. An enhancement of IX luminescence intensity and the narrow linewidth are observed in localized spots. 
\end{abstract}

\section{Introduction}

An indirect exciton (IX), also known as an interlayer exciton, is a bound pair of an electron and a hole confined in spatially separated layers. The spatial separation between the electron and hole layers allows achieving long IX lifetimes, orders of magnitude longer than lifetimes of direct excitons (DXs)~\cite{Lozovik1976}. Due to their long lifetimes, IXs can cool below the temperature of quantum degeneracy~\cite{Butov2001}. The realization of IX quantum Bose gases in GaAs heterostructures led to finding of many phenomena, including spontaneous coherence and condensation of IXs~\cite{High2012}, the spatially modulated exciton state~\cite{Butov2002, Alloing2014}, the commensurability effect of exciton density waves~\cite{Yang2015}, spin textures~\cite{High2013}, and the Pancharatnam-Berry phase and long-range coherent spin transport in the IX condensate~\cite{Leonard2018}.

Furthermore, an IX has a built-in electric dipole moment, $ed$ ($d$ is the separation between the electron and hole layers). As a result, IX energy, lifetime, and flux can be effectively controlled by voltage that is explored for the development of excitonic devices. In GaAs heterostructures, experimental proof-of-principle demonstrations were performed for excitonic ramps~\cite{Hagn1995, Gartner2006}, excitonic acoustic-wave~\cite{Lazic2014} and electrostatic~\cite{Winbow2011} conveyers, and excitonic transistors~\cite{High2008}.

However, the IX range of existence in GaAs heterostructures is limited to low temperatures due to low IX binding energies. Excitons exist in the temperature range roughly below $E_{\rm ex}/k_{\rm B}$ ($E_{\rm ex}$ is the exciton binding energy, $k_{\rm B}$ is the Boltzmann constant)~\cite{Chemla1984}. The IX binding energy in GaAs/AlGaAs heterostructures is typically $\sim 4$~meV~\cite{Sivalertporn2012}. The maximum $E_{\rm ex}$ in GaAs heterostructures is achieved in GaAs/AlAs coupled quantum wells (CQW) and is $\sim 10$~meV~\cite{Zrenner1992}. The temperature of quantum degeneracy, which can be achieved with increasing density before exciton dissociation to electron-hole plasma, also scales proportionally to $E_{\rm ex}$~\cite{Fogler2014}. In GaAs heterostructures, quantum degeneracy was achieved below few Kelvin~\cite{Butov2001} and the proof of principle for the operation of IX switching devices was demonstrated below $\sim 100$~K~\cite{Grosso2009}. IXs with high $E_{\rm ex}$ reaching $\sim 30$~meV are explored in ZnO and GaN heterostructures \cite{Lefebvre2004, Morhain2005, Fedichkin2016, Chiaruttini2019}.

Van der Waals heterostructures composed of atomically thin layers of TMD offer an opportunity to realize artificial materials with designable properties~\cite{Geim2013} and, in particular, allow the realization of excitons with remarkably high binding energies~\cite{Ye2014, Chernikov2014}. IXs in TMD heterostructures are characterized by binding energies exceeding 100~meV making them stable at room temperature~\cite{Fogler2014}. IXs were observed at room temperature in TMD heterostructures~\cite{Calman2018}. Due to the high IX binding energy, TMD heterostructures can form a material platform both for exploring high-temperature quantum Bose gases of IXs and for creating realistic excitonic devices.

\begin{figure}[H]
\begin{center}
\includegraphics[width=10cm]{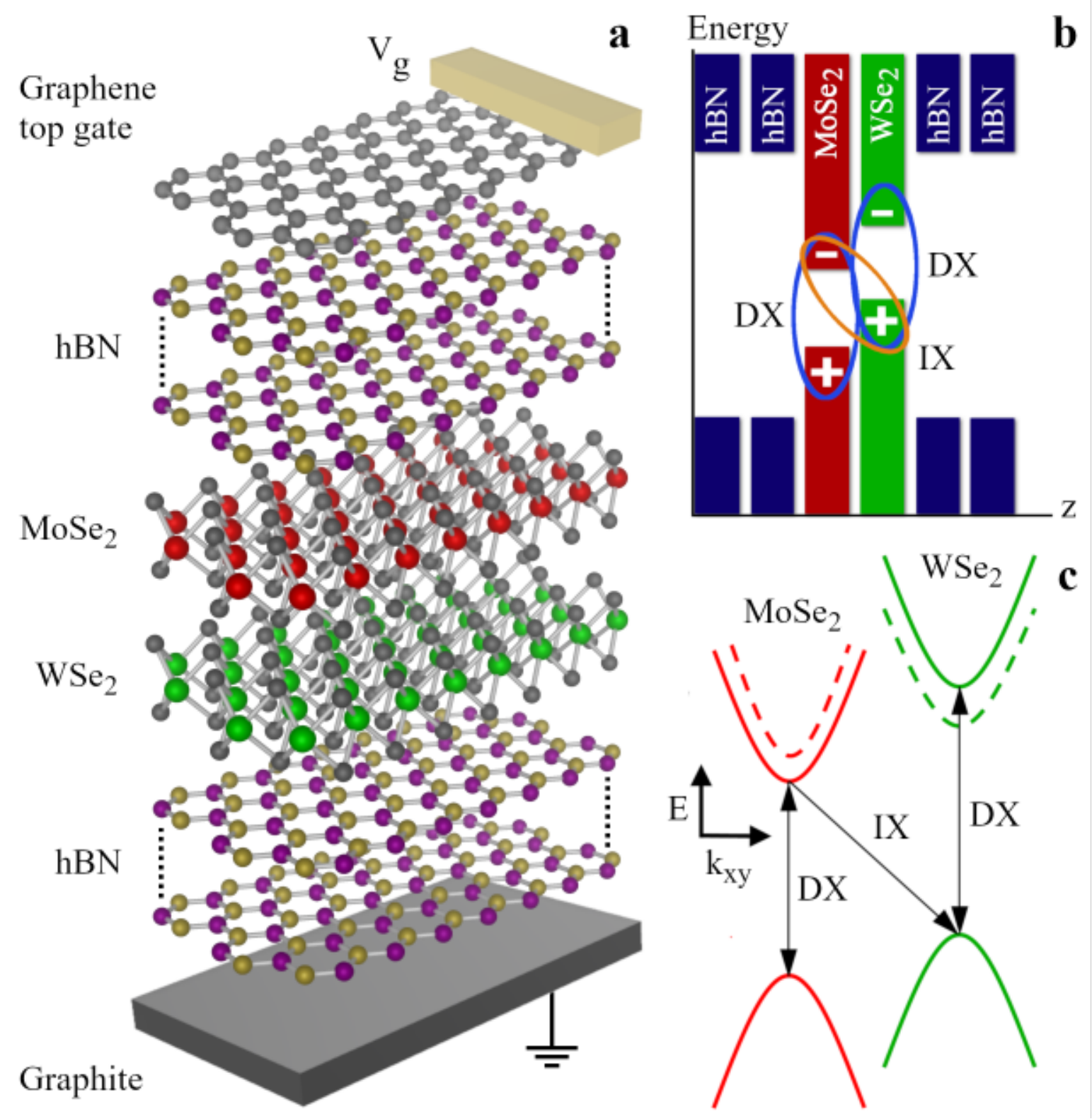}
\caption{\textbf{Van der Waals MoSe$_2$/WSe$_2$ heterostructure.} The heterostructure layer (a) and real space energy band (b) diagrams. The ovals indicate a direct exciton (DX) and an indirect exciton (IX) composed of an electron ($-$) and a hole ($+$). (c) Momentum space energy band diagram around the K point. Solid and dashed lines represent spin-up and spin-down bands. Optically active low-energy DX and IX states are indicated by arrows.}
\end{center}
\label{fig:spectra}
\end{figure}

IXs are instensively studied in optically excited van der Waals TMD heterostructures with coupled electron and hole layers~\cite{Fang2014, Hong2014, Rivera2015, Yu2015, Rivera2016, Miller2017, Nagler2017, Gao2018, Wang2018, Torun2018, Calman2018, Unuchek2018, Okada2018, Hanbicki2018, Jiang2018, Choi2018, Jauregui2018, Zhang2018, Ciarrocchi2019, Tran2019, Seyler2019, Jin2019, Alexeev2019, Jin2019a, Wu2017, Yu2017, Wu2018}. IXs can also appear in electron-electron (or hole-hole) bilayers in a collective electronic state in strong magnetic fields at the total Landau level filling factor 1. The latter was realized in GaAs heterostructures~\cite{Spielman2000, Kellogg2004, Tutuc2004, Eisenstein2004, Tiemann2008, Nandi2012} and in graphene--boron-nitride--graphene van der Waals heterostructures~\cite{Liu2017, Li2017}. 

\section{Results and discussion}

In this work, we present studies of IXs in MoSe$_2$/WSe$_2$ heterostructures. We report on the observation of charged IXs, i.e. indirect trions (IX$^{\rm T}$). The identification of indirect trions is based on the measured energy splitting and temperature dependence of IX and IX$^{\rm T}$ luminescence lines: The splitting corresponds to the binding energy for negative indirect trions in TMD heterostructures calculated in Ref.~\cite{Deilmann2018} and the temperature dependence follows the mass action law for the indirect trions. We also report on the realization of IXs with a luminescence linewidth reaching 4~meV at low temperatures, the lowest value reported so far for IXs in TMD heterostructures. An enhancement of IX luminescence intensity and the narrow linewidth are observed in localized spots. 

\begin{figure}
\begin{center}
\includegraphics[width=10cm]{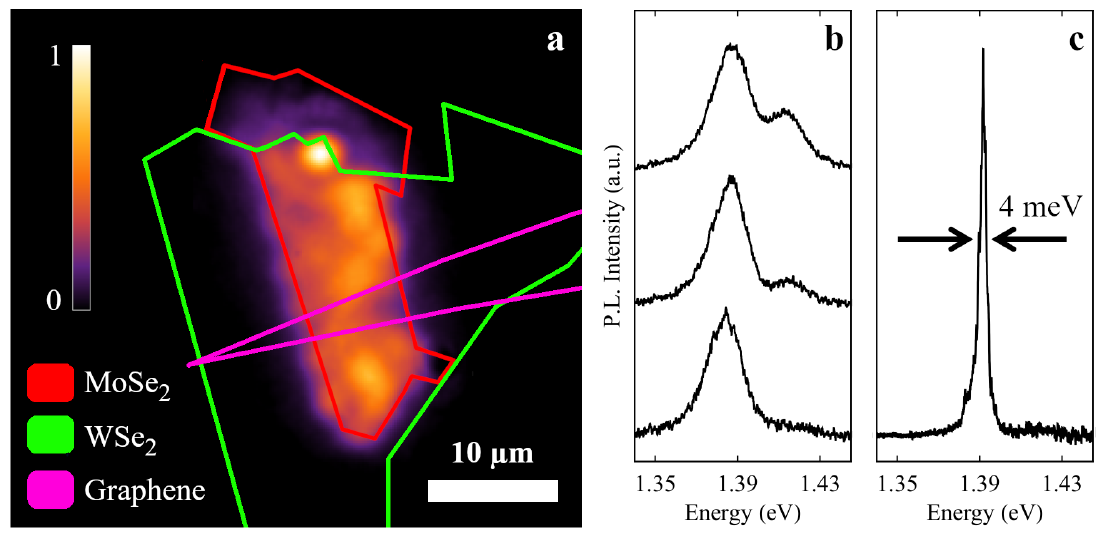}
\caption{\textbf{Spatially indirect, i.e., interlayer, luminescence in CQW flake and bright spot.} (a) $x$-$y$ map of indirect luminescence (spectral range 1.24--1.46~eV) in sample S. Indirect luminescence intensity is enhanced in a bright spot observed near the top of the CQW flake. The layer boundaries are shown. (b) The luminescence spectrum at the CQW flake in sample S at excitation power $P_{\rm ex} = 3.4$, 1, and 0.5~mW (top to bottom). (c) The luminescence spectrum at the bright spot in sample M at $P_{\rm ex} = 10$~$\mu$W. The laser excitation is defocused in (a) and focused at the flake center in sample S (b) and at the bright spot in sample M (c). $T=1.7$~K, $V_{\rm g}=0$.}
\end{center}
\label{fig:spectra}
\end{figure}

The MoSe$_2$/WSe$_2$ heterostructures were assembled by stacking mechanically exfoliated 2D crystals on a graphite substrate (Fig.~1a). The CQW is formed where the MoSe$_2$ and WSe$_2$ monolayers overlap. The MoSe$_2$ and WSe$_2$ monolayers are encapsulated by hexagonal boron nitride (hBN) serving as dielectric cladding layers. The real-space energy-band diagram is shown in Fig.~1b. IXs are formed from electrons and holes confined in adjacent monolayer MoSe$_2$ and WSe$_2$, respectively. These type-II MoSe$_2$/WSe$_2$ heterostructures with staggered band alignment are similar to AlAs/GaAs CQW where IXs are formed from electrons and holes confined in adjacent AlAs and GaAs layers, respectively~\cite{Zrenner1992, Grosso2009}. In the MoSe$_2$/WSe$_2$ heterostructures, due to the order of spin-up and spin-down states in valence and conduction bands (VB and CB) the lowest energy DX state is optically active in MoSe$_2$ and dark in WSe$_2$, and the lowest energy IX state is optically active (Fig.~1c)~\cite{Liu2013, Wang2015, Zhang2015, Zhou2017, Zhang2017}. We studied heterostructures manufactured in Manchester and San Diego (samples M and S). The order of MoSe$_2$ and WSe$_2$ layers is different in samples M and S to probe both configurations. Both samples show indirect trions and intensity enhancement in localized spots. 

Along most of the CQW heterostructure area, the IX luminescence intensity varies only slightly (Fig.~2a). We will refer to this CQW heterostructure area as the CQW flake. However, we observed bright spots, which exhibit enhanced IX luminescence in comparison to the surrounding regions of the CQW heterostructure [Fig.~2a and Figs.~S1 and S2 in Supplementary Information (SI)]. The CQW flakes and CQW bright spots show similar features of neutral and charged IX luminenscence and the data for both these regions are presented in this work.

In this paragraph, we outline phenomenological properties of the bright spots. We note that further details of their properties and their origin form the subject for future studies and do not affect the conclusions on neutral and charged IXs in this work. The enhancement of IX luminescence at the bright spots is localized within $\sim 2$~$\mu$m in sample S and within the length smaller than the 1~$\mu$m optics resolution in sample M. In contrast to IX luminescence, the intralayer DX luminescence varies only slightly along the CQW heterostructure and does not show an intensity enhancement in the bright spots [Figs.~S1 and S2 in SI]. The bright spots form naturally with no artificially designed IX confinement such as in electrostatic traps in GaAs heterostructures~\cite{Huber1998, Gorbunov2004, High2009nl, High2012nl, Schinner2013, Shilo2013, Mazuz-Harpaz2017}. The presence of the luminescence bright spot for IXs with a built-in electric dipole and its absence for DXs with no electric dipole suggests that the bright spots originate from an accidental IX trapping due to the background electrostatic potential in the heterostructures. The bright spot shows a narrow IX linewidth reaching 4~meV at the lowest excitation power tested (Figs.~2c and S3). 

\begin{figure}[H]
\begin{center}
\includegraphics[width=6cm]{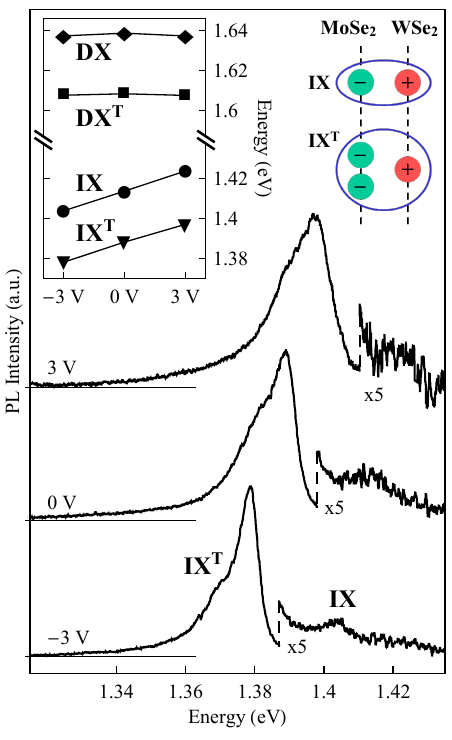}
\caption{\textbf{Gate voltage dependence.} Indirect luminescence spectra in MoSe$_2$/WSe$_2$ CQW at different gate voltages $V_{\rm g}$ at the bright spot in sample M. Left inset: Luminescence peak energy vs. $V_{\rm g}$. IX and IX$^{\rm T}$ are indirect exciton and trion, DX and DX$^{\rm T}$ are direct exciton and trion. Right inset: Schematic of IX and IX$^{\rm T}$. $P_{\rm ex}=1.25$~mW, $T=1.7$~K.}
\end{center}
\label{fig:Voltage}
\end{figure}

Two lines of spatially indirect luminescence are observed in the spectrum (Figs.~2b, 3). Due to the IX electric dipole moment, $ed$, the IX energy shifts in the voltage-induced electric field in the $z$ direction, $F_z$, by $\delta E = - edF_z$. The energy of two luminescence lines is controlled by voltage $V_{\rm g}$ applied between the graphene top gate and graphite back gate and creating the bias across the CQW structure (Fig.~3), indicating that both these lines correspond to spatially indirect luminescence. 

\begin{figure}
\begin{center}
\includegraphics[width=10cm]{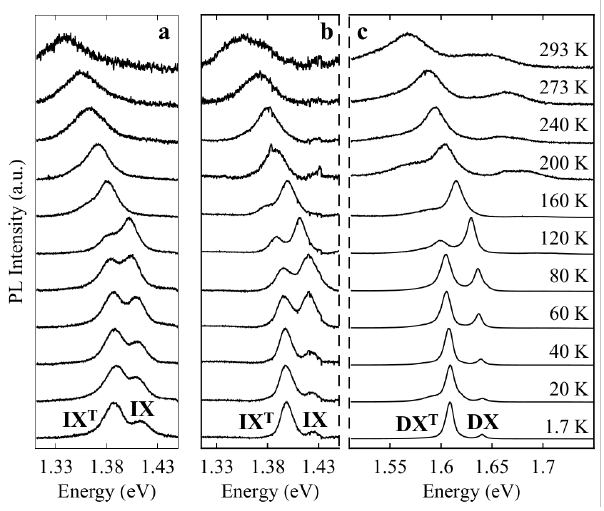}
\caption{\textbf{Temperature dependence.}  Spectra of spatially indirect (a,b) and direct (c) luminescence in MoSe$_2$/WSe$_2$ CQW at different temperatures at CQW flake in sample S (a) and bright spot in sample M (b,c). $P_{\rm ex} = 3.4$~(a) and 1.25 (b,c)~mW, $V_{\rm g}=0$.}
\end{center}
\label{fig:Temperature}
\end{figure}

The IX line splitting of 28 meV (sample S, Fig.~2b) [26~meV (sample M, Fig.~3)] is much smaller than the energy difference of the A and B excitons caused by the spin-orbit splitting of the WSe$_2$ VB~\cite{Zhu2011}. Therefore, both IX lines represent different species of A excitons. 

The measured energy splitting and temperature dependence of these two lines identify them as neutral and charged indirect excitons. The lower energy line corresponds to charged IXs, i.e., indirect trions (IX$^{\rm T}$), and the higher energy line to neutral IXs (Fig.~3 right inset).

The energy of the trion luminescence is determined by the difference between the initial state, trion, and final state, remaining electron (for negative trions). At low densities, the IX and IX$^{\rm T}$ luminescence energies should experience the same shift with voltage following the gap between the VB of WSe$_2$ and the CB of MoSe$_2$~\cite{Deilmann2018}, consistent with the experiment (Fig.~3). The splitting between the lines corresponds to the trion binding energy. Details are presented in SI. The experimentally found binding energy of the indirect trions of 26--28 meV is in agreement with the calculated binding energy of 28 meV for negative indirect trions in MoS$_2$/WS$_2$ heterostructures~\cite{Deilmann2018}.

Similarly, spatially direct neutral and charged excitons, DX and DX$^{\rm T}$, are observed for spatially direct, i.e., intralayer, luminescence (Fig.~4c). However, in contrast to IX and IX$^{\rm T}$, the peak energy of DX and DX$^{\rm T}$ practically does not change with voltage due to vanishing built-in dipole moment in the direction of applied electric field for direct excitons and trions (Fig.~3 left inset). 

The measured indirect trion binding energy of 26--28~meV is smaller than the direct trion binding energy of 32~meV (Fig.~4) due to the separation between the electron and hole layers, consistent with the theory of indirect trions in GaAs and TMD heterostructures~\cite{Witham2018, Bondarev2018, Deilmann2018}. DX and DX$^{\rm T}$ luminescence was studied earlier in monolayer MoSe$_2$~\cite{Mak2012, Ross2013, Berkelbach2013, Courtade2017, Zhou2017, Zhang2017}.

Further significant support for the asignment of the two lines of spatially indirect luminescence to neutral and charged indirect excitons comes from the temperature dependence: The luminescence intensity ratio of the lines IX$^{\rm T}$/IX decreases with increasing temperature (Fig.~4a,b, Fig.~5a red symbols, and Fig.~5b symbols) in agreement with the mass action law for the indirect trions (Fig.~5a red line and Fig.~5b lines). The relative intensity of the IX$^{\rm T}$ luminescence decreases with temperature due to the thermal dissociation of trions. The IX$^{\rm T}$ temperature dependence is similar to that for DX$^{\rm T}$ both in earlier studies of DX$^{\rm T}$ in MoSe$_2$ monolayers~\cite{Ross2013} and in this work (Figs.~4 and 5).

Solid lines in Fig.~5 present the simulated ratios of trion and exciton integrated luminescence intensities for the direct, DX$^{\rm T}$/DX, and the indirect, IX$^{\rm T}$/IX, cases. We simulated these ratios using their approximate proportionality to the densities of corresponding particles. The dependence of the densities on temperature is obtained from the mass action model~\cite{Ross2013}, details are presented in SI. In these simulations, the trion binding energy is taken from the measured line splitting. The simulations include two fitting parameters: the densities of background charge carriers $n_{\rm B}$ and photoexcited electron-hole pairs $n_{\rm P}$, their estimation is described in SI. The simulations give qualitatively similar results for various $n_{\rm P}$ and $n_{\rm B}$: At high temperatures, the ratio of trion and exciton densities $n_{\rm T} / n_{\rm X}$ increases with reducing temperature, however, at low temperatures, $n_{\rm T} / n_{\rm X}$ saturates (Figs.~5 and S8). This saturation is the key characteristic of trion luminescence. The origin of this saturation is in the finite number of background electrons that are invlolved in the trions. For the trions formed by binding of the background electrons with photoexcited excitons, at low temperatures, the trion density saturates at $n_{\rm B}$ and, in turn, the ratio $n_{\rm T} / n_{\rm X}$ asymptotically approaches $n_{\rm B} / (n_{\rm P} - n_{\rm B})$. The simulations are in agreement with the experimental data both for direct and indirect trions in the entire temperature range (Fig.~5).

As in the type-I MoS$_2$/hBN TMD heterostructure~\cite{Calman2018}, IXs are observed at room temperature in our type-II heterostructures (Fig.~4). The observed red shift of the lines with increasing temperature (Fig.~4) originates from the band gap reduction, which is typical for semiconductors, the TMDs included~\cite{Ross2013}.

\begin{figure}
\begin{center}
\includegraphics[width=10cm]{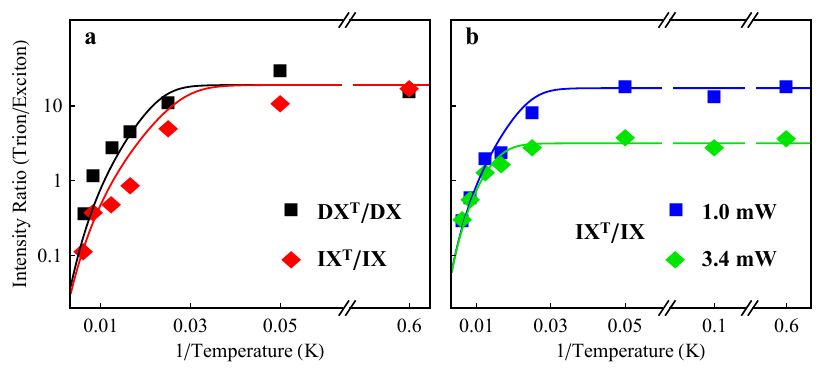}
\caption{\textbf{Temperature dependence.} Experimental (symbols) and simulated (lines) spectrally integrated luminescence intensity ratio IX$^{\rm T}$/IX (green, blue, red) and DX$^{\rm T}$/DX (black) vs. 1/temperature at the CQW bright spot in sample M (a) and the CQW flake in sample S (b). $P_{\rm ex} = 1.25$~mW (a), 1~mW [blue squares in (b)], and 3.4~mW [green diamonds in (b)], $V_{\rm g}=0$.}
\end{center}
\label{fig:Temperature2}
\end{figure}

The narrowest indirect luminescence linewidth is observed at the lowest temperature (Fig.~4) and smallest excitation power $P_{\rm ex}$ (Fig.~S3). The indirect luminescence broadens up to $\sim 40$~meV at room temperature (Fig.~4). With increasing $P_{\rm ex}$, the indirect luminescence broadens and shifts to higher energies (Figs.~2, S3). Similar line broadening and shift to higher energies were observed for IXs in GaAs heterostructures and described in terms of repulsive IX interaction~\cite{High2009nl}, which originates from the repulsion of oriented electric dipoles~\cite{Yoshioka1990, Zhu1995, Lozovik1996}. Increasing the density with $P_{\rm ex}$ leads to the enhancement of interaction in the system of indirect excitons and trions and, in turn, the enhancement of IX$^{\rm T}$ and IX energies.

Isolated IX$^{\rm T}$ have substantial binding energy at low separation between electron and hole layers~\cite{Witham2018, Bondarev2018, Deilmann2018}, relevant for the MoSe$_2$/WSe$_2$ heterostructure. However, the IX$^{\rm T}$ binding energy is smaller than the IX binding energy that stabilizes the neutral system of IXs against IX transformations to trions and charged particles. This suggests that most of IX$^{\rm T}$ form by binding of electrons and holes created by excitation to background charge carriers which are present in the heterostructure due to unintentional doping. Increasing $P_{\rm ex}$ leads to the enhancement of relative intensity of IX line, i.e. reduction of IX$^{\rm T}$/IX ratio, at low temperatures (Figs.~2b, 5b), consistent with the trion density saturation at $n_{\rm B}$ and, as a result, enhanced fraction of IXs with increased $n_{\rm P}$.

We also briefly discuss alternative intepretations for the two lines of spatially indirect luminescence. A splitting of IX or DX emission to two luminescence lines is a general phenomenon in two coupled TMD layers. Various interpretations based on the assignment of the lines to different states of neutral excitons were offered to explain this splitting: The interpretations in terms of (i) excitonic states split due to the CB K-valley spin splitting~\cite{Rivera2015}, (ii) excitonic states indirect in momentum space and split due to the valley energy difference~\cite{Miller2017, Okada2018} or spin-orbit coupling~\cite{Hanbicki2018}, and (iii) excitonic states in moir{\' e} superlattice~\cite{Zhang2018, Ciarrocchi2019, Seyler2019, Tran2019, Jin2019, Alexeev2019} following the theory of moir{\' e} IXs and DXs~\cite{Wu2017, Yu2017, Wu2018}.

However, interpretations based on different states of neutral excitons do not offer a good agreement with the experimental data in Fig.~5 and, in turn, a plausible explanation for the IX lines in the studied heterostructures. In particular, for different states of neutral excitons, the relative occupation of the lower-energy state and, as a result, the relative intensity of the lower-energy line should increase with lowering the temperature. This does not agree with the saturation of the relative intensity of the lower-energy line in the experiment (Fig.~5). The details are given in SI. In contrast, the theory of neutral excitons and trions is in agreement with the data (Fig.~5), indicating that the interpretation of the two lines based on neutral excitons and trions is more plausible.

In conclusion, we present studies of MoSe$_2$/WSe$_2$ heterostructures and report on two lines of spatially indirect luminescence whose energy splitting and temperature dependence identify them as neutral indirect excitons and charged indirect excitons, i.e. indirect trions.

\begin{acknowledgement}

We thank Michael Fogler for discussions. These studies were supported by DOE Office of Basic Energy Sciences under award DE-FG02-07ER46449 and by NSF Grant No.~1640173 and NERC, a subsidiary of SRC, through the SRC-NRI Center for Excitonic Devices.

\end{acknowledgement}

\begin{suppinfo}

\subsection{Map of indirect and direct luminescence}

The spatially indirect (interlayer) luminescence intensity is enhaced at the bright spots (sample S: Fig.~2a and Fig.~S1a; sample M: Fig.~S2a). Both in sample S and sample M, the bright spots are observed close to the flake boundary (Figs.~S1a and S2a). In contrast, the spatially direct (intralayer) luminescence intensity varies only slightly in the CQW heterostructures and does not show an intensity enhancement in the bright spots (sample S: Fig.~S1b,c; sample M: Fig.~S2b). 

\setcounter{figure}{0}
\begin{figure}[H]
\begin{center}
\includegraphics[width=15cm]{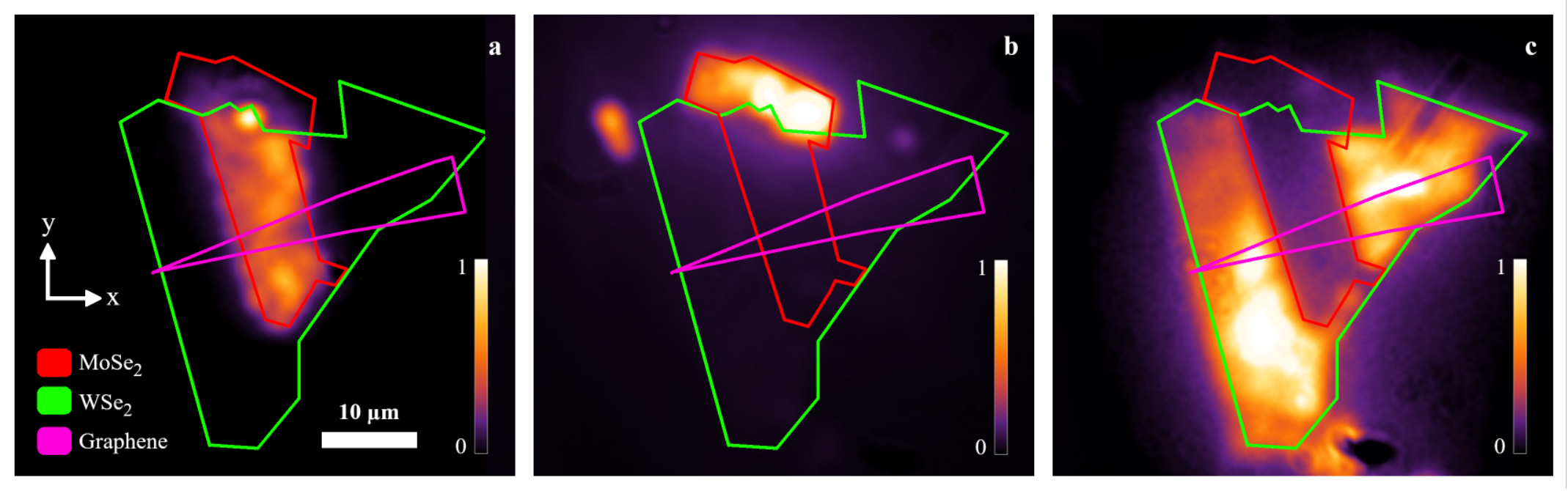}
\caption{\textbf{Map of indirect and direct luminescence in sample S.}  $x$-$y$ luminescence image of (a) the spatially indirect (interlayer) luminescence (measured in the spectral range 1.24--1.46~eV) and (b,c) the spatially direct (intralayer) luminescence in (b) MoSe$_2$ (measured in the spectral range 1.60--1.62eV~eV) and (c) WSe$_2$ (measured in the spectral range 1.66--1.69eV~eV) in sample S. The layer boundaries are shown. Laser excitation is defocused. $V_{\rm g}=0$, $T=1.7$~K.}
\end{center}
\label{fig:Voltagesup}
\end{figure}

\begin{figure}[H]
\begin{center}
\includegraphics[width=10cm]{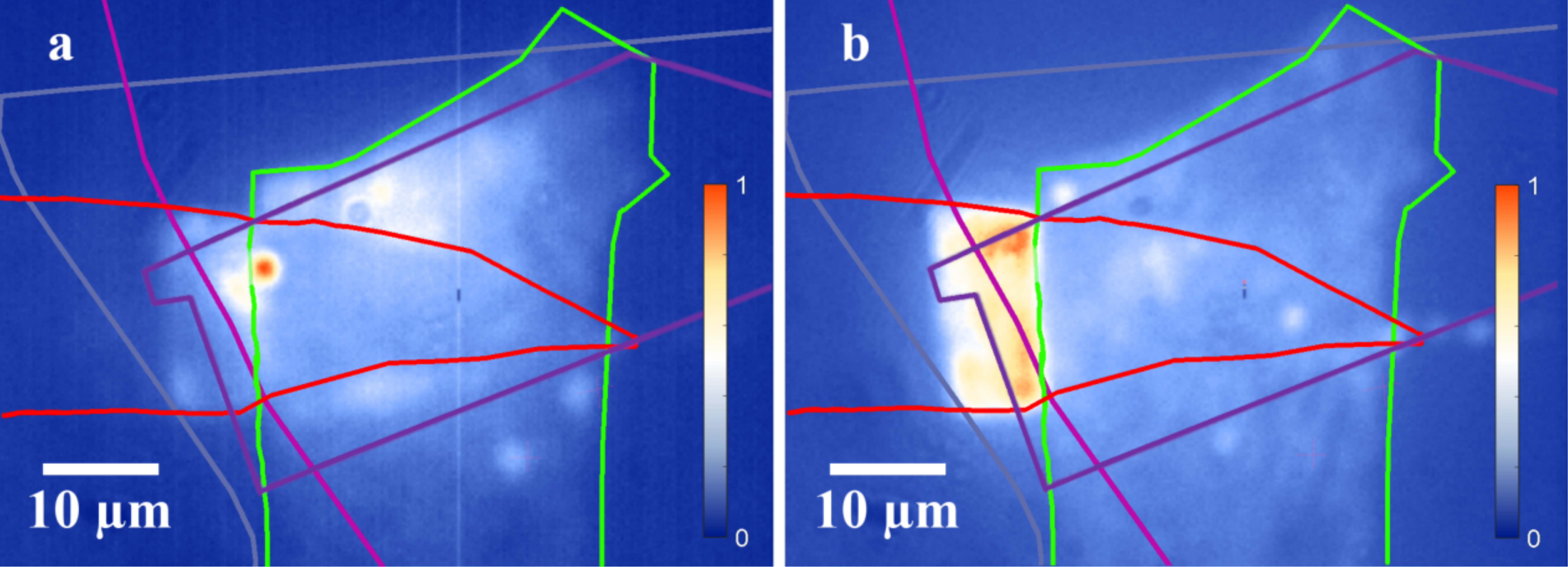}
\caption{\textbf{Map of indirect and direct luminescence in sample M.}  $x$-$y$ luminescence image of (a) the spatially indirect (interlayer) luminescence (measured in the spectral range 1.24--1.4~eV) and (b) the spatially direct (intralayer) luminescence (dominating in the spectral range 1.24--1.91~eV) in sample M. The layer boundaries are shown. Laser excitation is defocused. $V_{\rm g}=0$, $T=1.7$~K.}
\end{center}
\label{fig:Voltagesup}
\end{figure}

The bright spots show up to an order of magnitude enhancement of IX luminescence intensity in comparison to the surrounding region of the CQW heterostructure (Fig.~S2a). The IX luminescence in the bright spot is localized within $\sim 2$~$\mu$m in sample S and within the length smaller than the 1~$\mu$m resolution of the optical system used in the experiment in sample M.

\subsection{Excitation power dependence of indirect luminescence at the bright spot}

The narrowest indirect luminescence linewidth is observed at the smallest excitation power $P_{\rm ex}$ in the experiment. With increasing $P_{\rm ex}$, the indirect luminescence broadens (Fig.~S3a,d) and shifts to higher energies (Fig.~S3a,b).

\begin{figure}[H]
\begin{center}
\includegraphics[width=7cm]{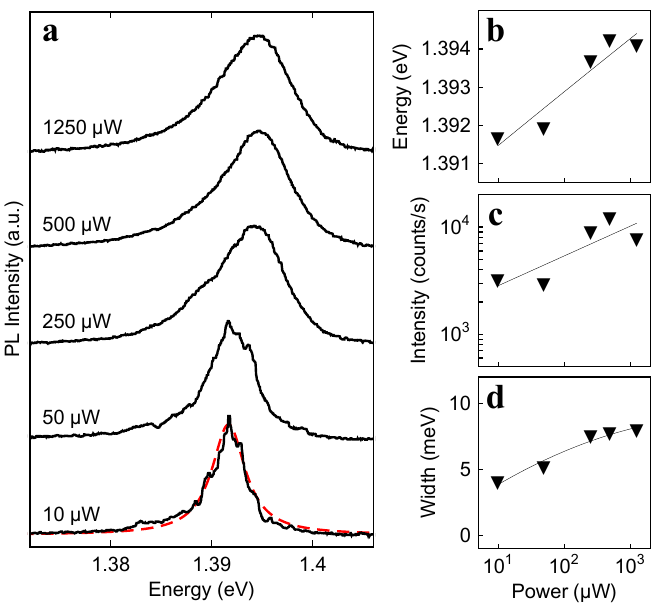}
\caption{\textbf{Excitation power dependence of indirect luminescence at the bright spot in sample M}. (a) Indirect luminescence spectra at different excitation powers $P_{\rm ex}$. (b-d) The peak energy (b), intensity (c), and linewidth (d) of the luminescence line in (a) vs. $P_{\rm ex}$. These parameters are extracted from Lorentzian fits to the luminescence lines, an example is shown for the 10~$\mu$W spectrum by a red dashed line in (a). The curves are guides to the eye. $T = 1.7$~K, $V_{\rm g}=0$.}
\end{center}
\label{fig:Voltagesup}
\end{figure}

\subsection{Voltage dependence of luminescence at the CQW flake}

The IX energy at the CQW flake is controlled by voltage $V_{\rm g}$ (Figs.~3, S4, S5). For sample S, the bias across the CQW is applied by a narrow graphene stripe on the top of the heterostructure (Fig.~2a) and the voltage dependence presented in Fig.~S4 is measured at the graphene stripe location. For sample M, the graphene layer covers the entire CQW flake and the voltage dependence in Fig.~S5 is measured at the flake center. With increasing $V_{\rm g}$, the IX energy increases in sample M (Fig.~3 and S5) and reduces in sample S (Fig.~S4), in agreement with the different order of MoSe$_2$ and WSe$_2$ layers in samples M and S. The neutral and charged indirect exciton peaks, IX and IX$^{\rm T}$, are not resolved at the CQW flake in sample M due to larger luminescence linewidth in this region of the sample. The energy shifts with voltage in the CQW flake (Fig.~S5) and bright spot (Fig.~3) are roughly the same, indicating that in both these regions the indirect luminescence lines correspond to IXs with the same $d$. 

\begin{figure}[H]
\begin{center}
\includegraphics[width=5cm]{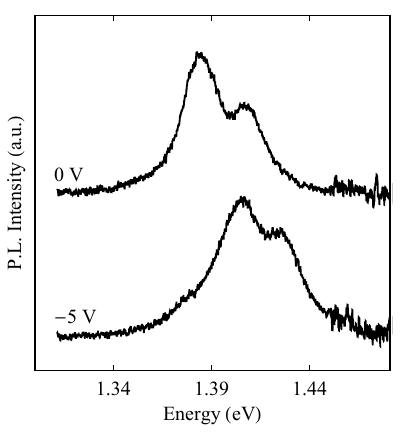}
\caption{\textbf{Voltage dependence of luminescence at the CQW flake in sample S.} Indirect luminescence spectra at the CQW flake [at (0, $-13$~$\mu$m), 
the coordinate center is at the bright spot] at different gate voltages $V_{\rm g}$. $P_{\rm ex}=3.4$~mW. $T = 1.7$~K.}
\end{center}
\label{fig:Voltagesup}
\end{figure}

\begin{figure}[H]
\begin{center}
\includegraphics[width=5cm]{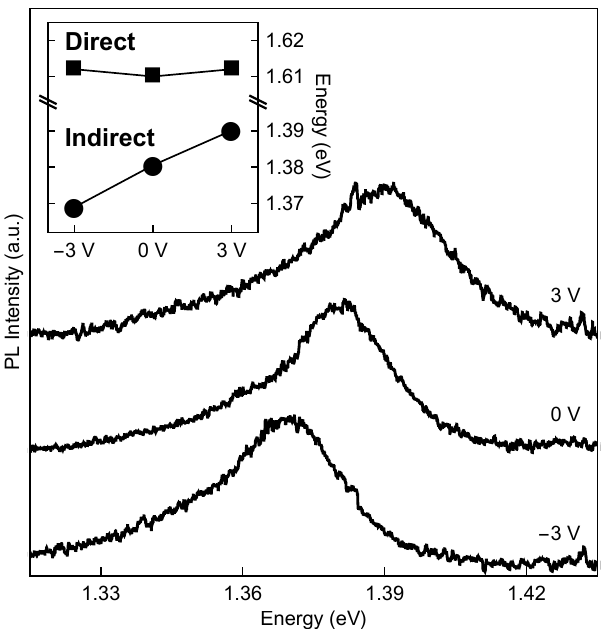}
\caption{\textbf{Voltage dependence of luminescence at the CQW flake in sample M.} Indirect luminescence spectra at the CQW flake [at (15~$\mu$m, $-4$~$\mu$m), the coordinate center is at the bright spot] at different gate voltages $V_{\rm g}$. The neutral and charged indirect exciton peaks, IX and IX$^{\rm T}$, are not resolved due to larger luminescence linewidth in this region of the sample. The inset shows the peak energy of direct and indirect luminescence lines vs. $V_{\rm g}$. $P_{\rm ex}=1.25$~mW, $T=1.7$~K.}
\end{center}
\label{fig:Voltagesup}
\end{figure}

\subsection{Temperature dependence of luminescence in sample S}

Temperature dependences of luminescence at the CQW flake (Figs.~4a and 5b repeated in Figs.~S6b and 6e) and at the bright spot (Fig.~S6d,f) in sample S are similar. Similar temperature dependences are also observed for low (Fig.~S6a,c) and high (Fig.~S6b,d) excitation powers. With reducing temperature, the luminescence intensity ratio of the lines IX$^{\rm T}$/IX increases at high temperatures and saturates at low temperatures (Fig.~S6a-d and symbols at Fig.~6e,f) in agreement with the mass action law for the indirect trions (blue and green lines in Fig.~S6e,f).

\begin{figure}[H]
\begin{center}
\includegraphics[width=14cm]{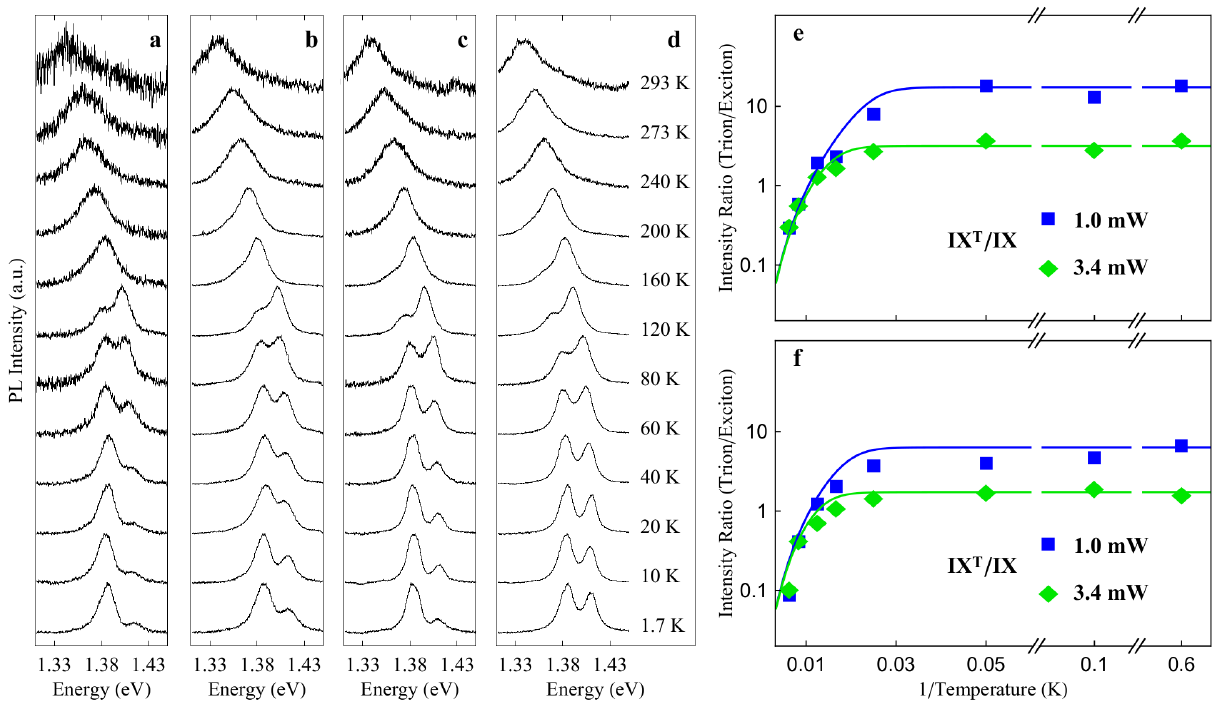}
\caption{\textbf{Temperature dependence of indirect luminescence in sample S}. (a-d) Indirect luminescence spectra at the CQW flake [at (0, $-7$~$\mu$m)] (a,b) and at the CQW bright spot (c,d) at different temperatures. Excitation power $P_{\rm ex} = 1$~mW (a,c) and 3.4~mW (b,d). (e,f) Experimental (symbols) and simulated (lines) spectrally integrated luminescence intensity ratio IX$^{\rm T}$/IX vs. 1/temperature at the CQW flake (e) and at the CQW bright spot (f). $P_{\rm ex} = 1$~mW (blue squares) and 3.4 mW (green diamonds), $V_{\rm g}=0$.}
\end{center}
\label{fig:Voltagesup}
\end{figure}

\subsection{Temperature dependence of luminescence at the CQW flake in sample M}

Figure~S7 shows indirect and direct luminescence spectra at the CQW flake in sample M at different temperatures. The fit to indirect spectra in Fig.~S7a and direct spectra in Fig.~S7b is shown in Fig.~S7c and Fig.~S7d, respectively. In contrast to the bright spot (Fig.~4), the neutral and charged indirect exciton peaks, IX and IX$^{\rm T}$, are not resolved in the CQW flake in sample M due to larger luminescence linewidth in this region (Fig.~S7a). However, the fit to spectra (Fig.~S7c) suggests that the temperature dependence of indirect luminescence at the CQW flake in sample M is similar to that at the bright spot in sample M (Fig.~4b), at the bright spot in sample S (Fig.~S6), and at the CQW flake in sample S (Fig.~4a): At low temperatures, the indirect spectra are dominated by the IX$^{\rm T}$ luminescence (green Lorentzians in Fig.~S7c), while the relative intensity of the high-energy IX luminescence (blue Lorentzian in Fig.~S7c) increases with temperature, consistent with the thermal dissociation of IX$^{\rm T}$.

Both MoSe$_2$ and WSe$_2$ contribute to direct luminescence in the CQW heterostructure (Fig.~S7b). As outlined in the main text, the lowest energy DX state is optically active (bright) in MoSe$_2$ and dark in WSe$_2$ (Fig.~1c)~\cite{Jiang2018, Liu2013, Wang2015, Zhang2015, Zhou2017, Zhang2017}. As a result, at low temperatures, the direct luminescence in the CQW heterostructure is dominated by MoSe$_2$. The intensity of direct luminescence of WSe$_2$ grows with increasing temperature due to thermal occupation of higher-energy conduction band states involved in the optically active DXs (Fig.~1c).

At the lowest tested temperatures, the direct spectra are dominated by MoSe$_2$ DX$^{\rm T}$ luminescence line (green Lorentzians in Fig.~S7d). The relative intensity of the higher-energy MoSe$_2$ DX luminescence (blue Lorentzians in Fig.~S7d) increases with temperature, consistent with the thermal dissociation of MoSe$_2$ DX$^{\rm T}$. The low-energy shoulders (magenta Lorentzians in Fig.~S7c,d) are tentatively attributed to localized states. 

WSe$_2$ DX$^{\rm T}$ luminescence line (cyan Lorentzians in Fig.~S7d) and WSe$_2$ DX luminescence line (orange Lorentzians in Fig.~S7d) appear in the spectrum around 60~K. The relative intensity of the higher-energy WSe$_2$ DX luminescence increases with temperature, consistent with the thermal dissociation of WSe$_2$ DX$^{\rm T}$. The red shift of the lines with increasing temperature (Fig.~S7e) originates from the band gap reduction, which is typical in semiconductors, the TMDs included~\cite{Ross2013}.

\begin{figure}[H]
\begin{center}
\includegraphics[width=15cm]{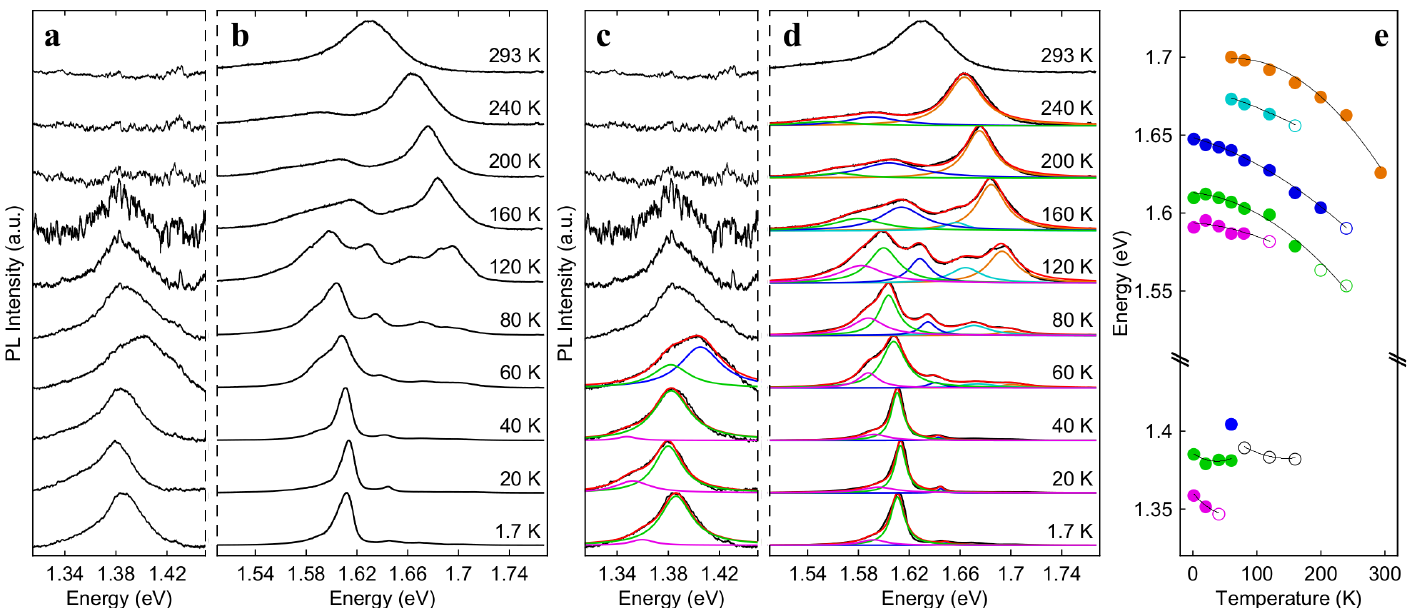}
\caption{\textbf{Temperature dependence of luminescence at the CQW flake in sample M}.  (a) Indirect and (b) direct luminescence spectra at the CQW flake [at (15~$\mu$m, $-4$~$\mu$m)] at different temperatures. The fit to spectra in (a) and (b) is shown in (c) and (d), respectively. The spectra are shown by black lines. The fit Lorentzians are shown by green (IX$^{\rm T}$ and MoSe$_2$ DX$^{\rm T}$), blue (IX and MoSe$_2$ DX), cyan (WSe$_2$ DX$^{\rm T}$), orange (WSe$_2$ DX), and magenta (localized states) lines, their sums by red lines. (e) The peak energy of luminescence lines vs. temperature. The color of dots in (e) corresponds to the color of corresponding luminescence lines in (c,d). The curves are guides to the eye. $P_{\rm ex}=1.25$~mW, $V_{\rm g}=0$.}
\end{center}
\label{fig:Voltagesup}
\end{figure}

\subsection{Simulations}

The simulations set off with determining the densities of excitons and trions from the mass action model following~\cite{Ross2013}. The mass of an exciton is $m_X = m_e+m_h$ and that of a trion is $m_T = 2 m_e+m_h$. The densities of excitons $n_X$, trions $n_T$, and free electrons $n_e$ are determined from conditions
\begin{eqnarray}
n_P & = & n_X + n_T, \\
n_B & = & n_T + n_e, \\
n_X n_e/n_T & = & A k_BT  exp \left( -\frac{E_T}{k_BT} \right),
\end{eqnarray}
where $k_B$ is the Boltzmann constant, $T$ is the ambient temperature, $n_B$ is the density of background electron doping, $n_P$ is the density of photoexcited electron-hole pairs, $E_T$ is the trion binding energy, and $A = (4 m_e m_X)/(\pi \hbar^2 m_T)$. The trion binding energy is taken from the measured line splitting: $E_T = 28$~meV for indirect trion in sample S and 26~meV for indirect trion in sample M and $E_T = 32$~meV for direct trion in MoSe$_2$ (Fig.~4). This model gives the ratio of exciton and trion densities presented in Figs.~5 and S6 by black (for direct exciton and trion) and red, green, and blue (for indirect exciton and trion) lines.

The simulations give qualitatively similar results for various $n_P$ and $n_B$: At high temperatures, the ratio $n_T/n_X$ increases with reducing temperature, however, at low temperatures,  $n_T/n_X$ saturates (Fig.~S8). This saturation is the key characteristic of trion luminescence as outlined in the main text. 

\begin{figure}[H]
\begin{center}
\includegraphics[width=10.5cm]{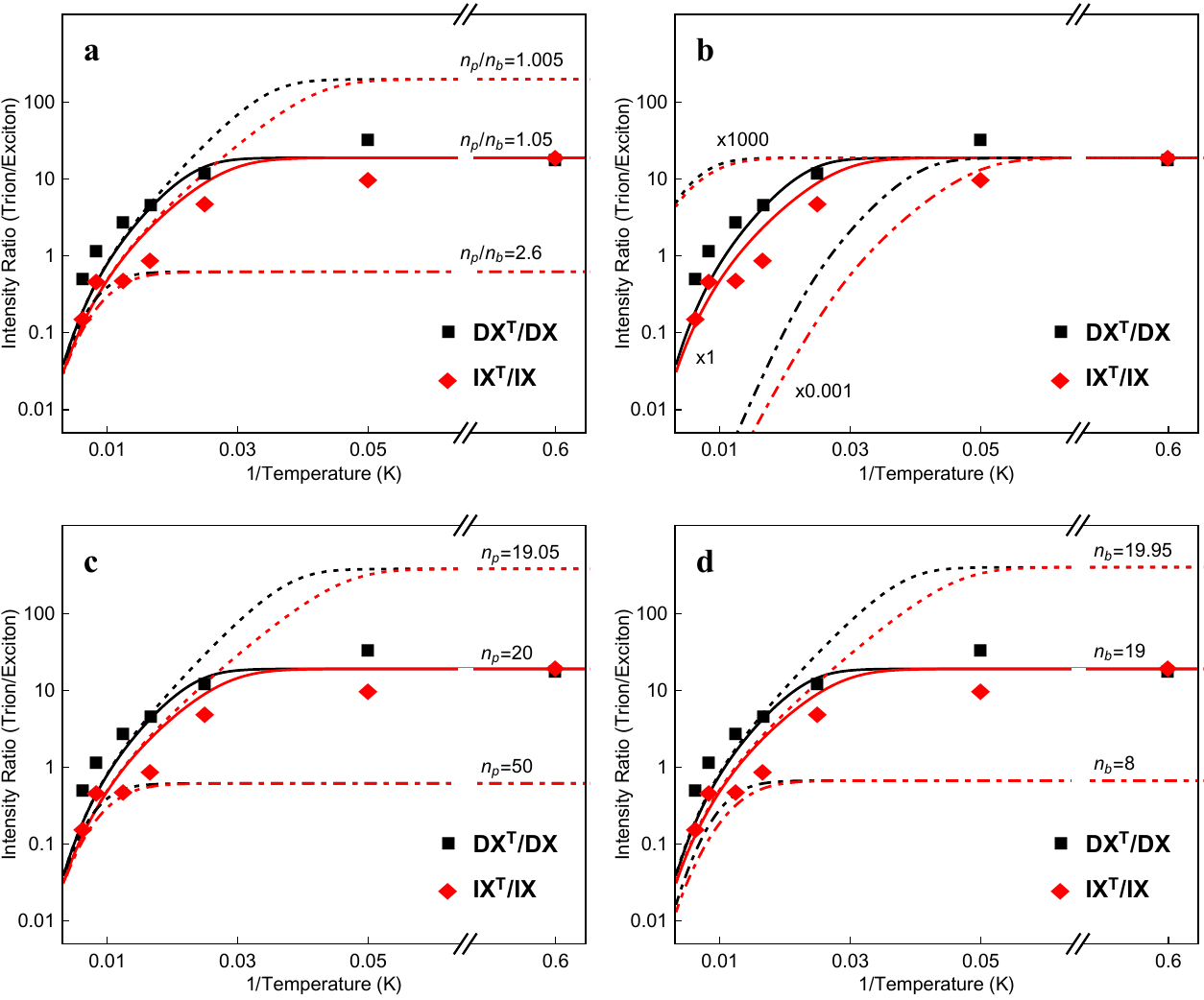}
\caption{\textbf{Estimating $n_P$ and $n_B$}. The measured (symbols) and simulated (lines) spectrally integrated luminescence intensity ratio IX$^{\rm T}$/IX (red) and DX$^{\rm T}$/DX (black) vs. 1/temperature (a) for fixed $n_B=1.9 \cdot 10^{11}$~cm$^{-2}$ fitted to the high-temperature data and different $n_P / n_B$, (b) for fixed $n_P/n_B = 1.05$ fitted to the low-temperature data and different $n_B$ (solid lines for $n_B=1.9 \cdot 10^{11}$~cm$^{-2}$, dotted and dashed-dotted lines for 1000 times higher and lower $n_B$, respectively), (c) for fixed $n_P=2 \cdot 10^{11}$~cm$^{-2}$ and different $n_B$, (d) for fixed $n_B=1.9 \cdot 10^{11}$~cm$^{-2}$ and different $n_P$. The densities indicated in the figure are in $10^{10}$~cm$^{-2}$.}
\end{center}
 \label{fig:Voltagesup}
\end{figure}

An estimate for $n_P$ and $n_B$ can be obtained by treating them as fitting parameters (Fig.~S8). The procedure of estimating $n_P$ and $n_B$ is simplified by separate fitting of the ratio $n_P/n_B$ and then $n_B$ (or $n_P$) to the low- and high-temperature data, respectively, as described below. In the limit of vanishing temperatures, equations (1)--(3) give the expected result: the trion density saturates at $n_B$ and, in turn, the ratio $n_T/n_X$ asymptotically approaches $n_B/(n_P - n_B) = 1/(n_P/n_B - 1)$. Therefore fitting the low-temperature data gives an estimate for $n_P / n_B$ (Fig.~S8a). Then $n_B$ and, in turn, $n_P$ can be estimated by fitting the high-temperature data (Fig.~S8b). The obtained fitting parameters for sample M are $n_B=1.9 \cdot 10^{11}$~cm$^{-2}$ and $n_P=2 \cdot 10^{11}$~cm$^{-2}$. Figures~S8c and S8d show the variation of $n_T/n_X$ with $n_P$ and $n_B$, respectively. 

The measured temperature dependence of indirect exciton and trion luminescence in the MoSe$_2$/WSe$_2$ CQW heterostructure as well as direct exciton and trion luminescence in MoSe$_2$ monolayer and in WSe$_2$ monolayer was compared with simulations based on the mass action model outlined above. The spectra of indirect luminescence were measured at the bright spot in MoSe$_2$/WSe$_2$ CQW heterostructure in sample M (Fig.~S9a). The spectra of direct luminescence in MoSe$_2$ monolayer (Fig.~S9b) and in WSe$_2$ monolayer (Fig.~S9c) were measured outside the CQW heterostructure region, namely in the monolayer MoSe$_2$ and in the monolayer WSe$_2$ regions of sample M. 

\begin{figure}[H]
\begin{center}
\includegraphics[width=12cm]{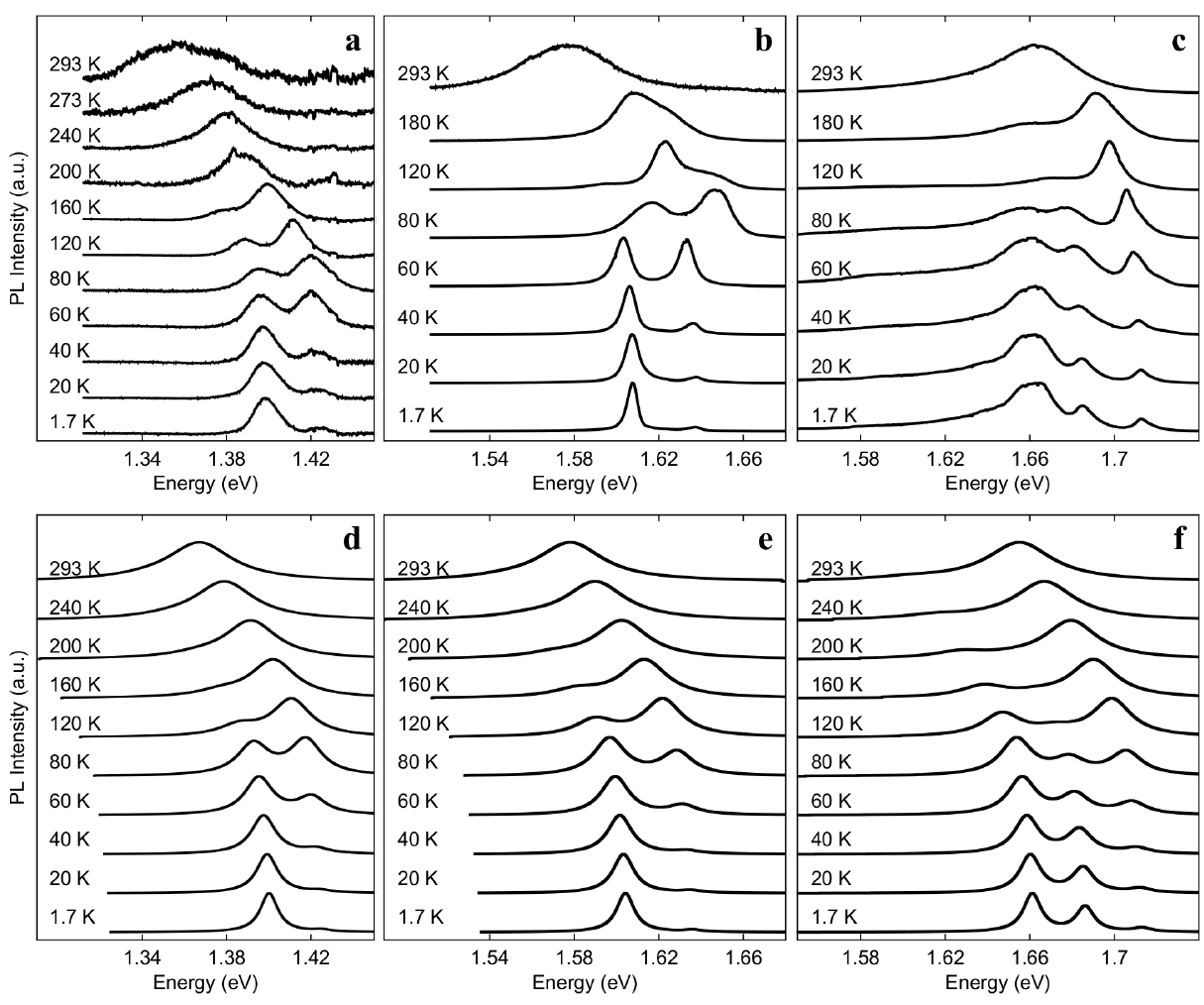}
\caption{\textbf{Measured and simulated temperature dependence of luminescence at the bright spot in MoSe$_2$/WSe$_2$ CQW heterostructure and in MoSe$_2$ and WSe$_2$ monolayers in sample M.} Measured spectra of (a) indirect luminescence in the bright spot in MoSe$_2$/WSe$_2$ CQW heterostructure, (b) direct luminescence in MoSe$_2$ monolayer, and (c) direct luminescence in WSe$_2$ monolayer at different temperatures. $P_{\rm ex}=1.25$~mW, $V_{\rm g}=0$. Simulated luminescence of (d) IX and IX$^{\rm T}$ in MoSe$_2$/WSe$_2$ CQW heterostructure, (e) DX and DX$^{\rm T}$ in MoSe$_2$ monolayer, and (f) DX and DX$^{\rm T}$ in WSe$_2$ monolayer.}
\end{center}
 \label{fig:Voltagesup}
\end{figure}

Two species of both excitons and trions (bright and dark) were observed in direct luminescence of monolayer WSe$_2$~\cite{Zhou2017, Zhang2017}. In the monolayer WSe$_2$ studied here, three luminescence lines can be seen (Fig.~S9c). The simulations of WSe$_2$ luminescence (Fig.~S9f) include direct exciton and two species of direct trions with the binding energies taken from the measured line splitting $E_{T1} = 27$~meV and $E_{T2} = 52$~meV (Fig.~S9c).

Figures~S9d-f show luminescence spectra simulated using the following approximation. The ratio of exciton and trion luminescence intensities is taken equal to the ratio of exciton and trion densities. The estimation of the latter is described above. A phenomenological dependence of the bandgap on temperature is taken from the experiment (Fig.~S9a-c) $\Delta E_g [{\rm meV}] = - 0.05 T - 6 \cdot 10^{-4} T^2$, where temperature is in units of Kelvin. The peak energies of exciton luminescence, i.e., the bandgap minus the exciton binding energy, at zero temperature are taken from the experiment (Fig.~S9a-c)
$E_g - E_X = {1425}$~meV (IX in CQW), 
$E_g - E_X = {1636}$~meV (DX in MoSe$_2$),
$E_g - E_X = {1713}$~meV (DX in WSe$_2$). 
The ratio of trion 1 and trion 2 densities in WSe$_2$ is approximated by the Maxwell-Boltzmann distribution. We do not specifically consider the effects of interaction, disorder, recoil, or light cone on the luminescence lineshape. Instead, we incorporate a phenomenological Lorentzian broadening dependent on temperature as $\Delta [{\rm meV}] = 5 + 0.05 T$ for the luminescense lines. The simulated spectra (Fig.~S9d-f) illustrate the major features observed in the experiments (Fig.~S9a-c).

\subsection{Two indirect luminescence lines: Interpretation in terms of indirect trion IX$^{\rm T}$ and neutral indirect exciton IX \\ vs. interpretation in terms of two different states of neutral IXs}

Figure~S10 reproduces the measured (symbols) and simulated (solid lines) spectrally integrated line luminescence intensity ratio IX$^{\rm T}$/IX (red) and DX$^{\rm T}$/DX (black) vs. $1/{\rm temperature}$ (same data as in Fig.~5a). The simulation of the indirect trion IX$^{\rm T}$ to indirect neutral exciton IX line intensity ratio (solid red line) is in agreement with the experiment (red symbols) as outlined above. 

We also consider alternative interpretations for the two indirect luminescence lines. In Fig.~S10, the measured temperature dependence of the line intensity ratio (Fig.~S10 symbols) is also compared with the ratio expected for different states of neutral IXs. As for the model of a trion and a neutral exciton outlined above, we consider the states at thermal equilibrium and obeying the Maxwell-Boltzmann distribution. Thermal equilibrium is facilitated by the long IX lifetimes.

The exciton luminescence intensity is given by $n_X  /\tau_r$ so the luminescence intensity ratio for two states of neutral IXs is given by $n_{X1} / n_{X2} \cdot \tau_{r2} / \tau_{r1} = n_{X1}/n_{X2} \cdot R$, where $R$ is the ratio of radiative lifetimes $\tau_{r2}$ and $\tau_{r1}$ of the two excitonic transitions. For the splitting between two states of neutral IXs $\Delta E = 26$~meV corresponding to the measured splitting between the lines, their luminescence intensity ratio is proportional to their occupation ratio $\exp(\Delta E / k_{\rm B}T)$ shown in Fig.~S10 by red dashed line. Treating $R$ as an adjustable parameter for achieving as good as possible agreement with the experiment allows shifting the red dashed line vertically keeping its slope unchanged in Fig.~S10. For the red dashed line in Fig.~S10, $R \sim 150$ (with the higher energy exciton state having 150 times shorter radiative lifetime than the lower energy exciton state) to fit the experimental point at $T = 60$~K, this provides a better fit to the experiment, however, assuming other $R$ does not change the discussion below. 
[For a model based on two states of neutral IXs, an explanation of the higher intensity of the higher energy exciton line observed at $T > 60$~K (Figs.~4 and S6) requires a significantly higher oscillator strength for the higher energy exciton state.]
Figure~S10 shows that $\exp(\Delta E / k_{\rm B}T)$ (red dashed line) does not agree with the experimental data. Since varying the ratio $R$ of radiative lifetimes of the two excitonic transitions only moves the dashed line vertically in Fig.~S10, no agreement is achieved for any $R$. For instance, in the range $T = 20 - 160$~K the measured intensity ratio drops by $\sim 50$ times while the intensity ratio for two states of neutral IXs $\exp(\Delta E / k_{\rm B}T)$ drops by $\sim 5 \times 10^5$ times (Fig.~S10). This large discrepancy, by orders of magnitude, indicates that interpretations of the two indirect luminescence lines based on two different states of neutral IXs are less plausible than the interpretation based on neutral exciton IX and trion IX$^{\rm T}$, which is in agreement with the data.

\begin{figure}[H]
\begin{center}
\includegraphics[width=5.5cm]{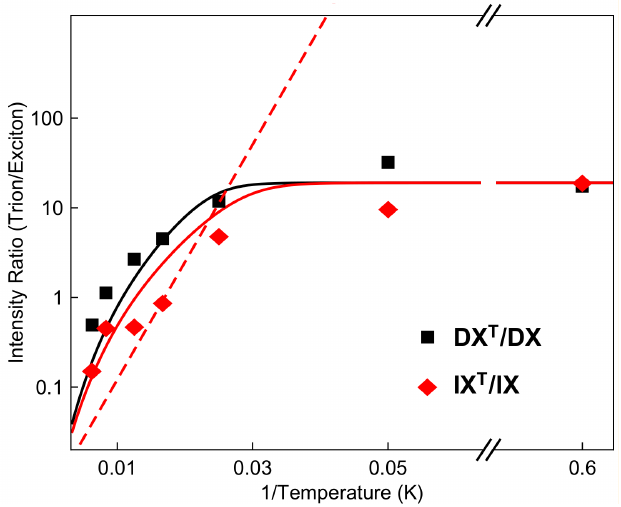}
\caption{\textbf{Temperature dependence of luminescence}. The measured (symbols) and simulated (solid lines) spectrally integrated luminescence intensity ratio IX$^{\rm T}$/IX (red) and DX$^{\rm T}$/DX (black) vs. $1/{\rm temperature}$  (same data as in Fig.~5a). The simulation of the indirect trion IX$^{\rm T}$ to indirect neutral exciton IX line intensity ratio (solid red line) is in agreement with the experiment (red symbols). Exponential ratio $\exp(\Delta E / k_{\rm B}T)$ for two states of neutral IXs separated by $\Delta E = 26$~meV (red dashed line) does not offer a good agreement with the experiment.}
\end{center}
\label{fig:Voltagesup}
\end{figure}

\subsection{Power dependence of luminescence at the CQW flake in sample M}

As at the bright spot (Fig.~S3), the indirect luminescence outside the bright spot at the CQW flake shifts to higher energies with increasing excitation power $P_{\rm ex}$ (Fig.~S11). No such energy enhancement is observed for direct luminescence neither in the bright spot no outside the bright spot. 

\begin{figure}[H]
\begin{center}
\includegraphics[width=11cm]{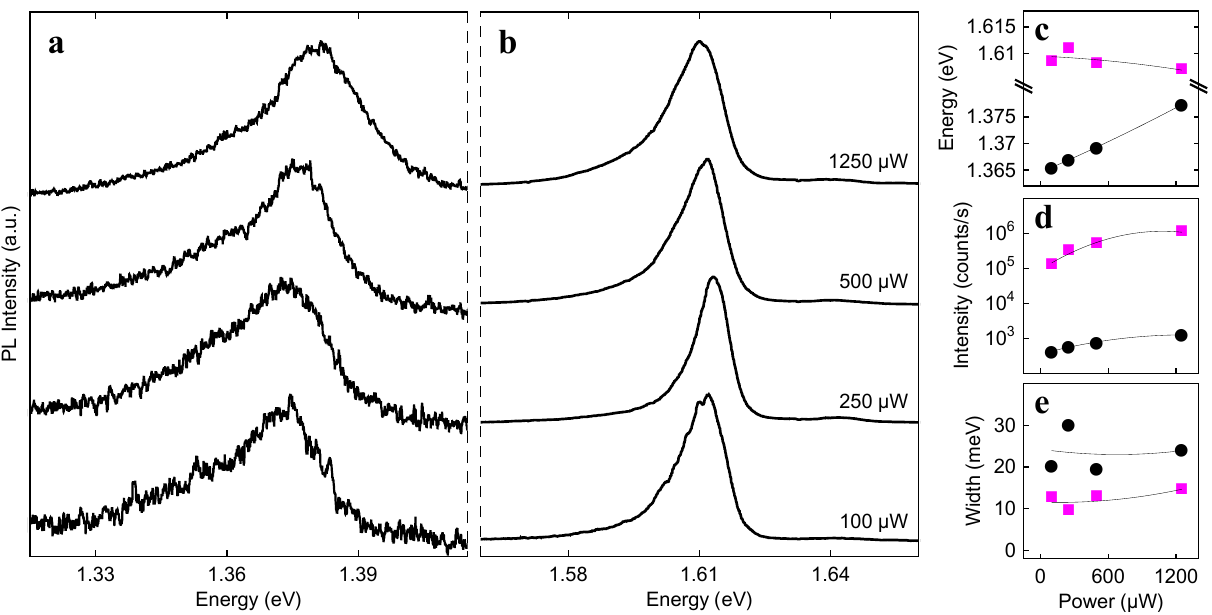}
\caption{\textbf{Power dependence of luminescence at the CQW flake in sample M}. (a) Indirect and (b) direct luminescence spectra at the CQW flake [at (15~$\mu$m, $-4$~$\mu$m)] in sample M at different excitation powers $P_{\rm ex}$. The energy (c), intensity (d), and linewidth (e) of the indirect (black circles) and direct (magenta squares) luminescence lines shown in (a) and (b) vs. $P_{\rm ex}$. $V_{\rm g}=0$, $T=1.7$~K.}
\end{center}
\label{fig:Voltagesup}
\end{figure}

\subsection{Shift of IX$^{\rm T}$ and IX luminescence lines with voltage} 

The energy of the trion luminescence is determined by the difference between the initial state, trion, and final state, remaining electron (for negative trions). At low densities, the IX and IX$^{\rm T}$ luminescence energies are given by $E_0 - edF_z - E_{\rm X}$ and $E_0 - edF_z - E_{\rm X} - E_{\rm T}$, respectively, and IX and IX$^{\rm T}$ luminescence lines should experience the same shift with voltage following the gap between the VB of WSe$_2$ and the CB of MoSe$_2$ \cite{Deilmann2018}, consistent with the experiment (Fig.~3) ($E_0 - edF_z$ is the gap between the VB of WSe$_2$ and the CB of MoSe$_2$, $E_{\rm X}$ and $E_{\rm  T}$ are IX and IX$^{\rm T}$ binding energies, respectively). While the trion has an addition electron, the energy of this electron in the electric field $F_z$ enters both to the initial and final state and, as a result, does not affect the trion luminescence energy. 

At high densities, the shifts may become different due to collective phenomena. For instance, for the trion luminescence, the final state of the remaining electron is affected by the electron Fermi energy, and, if the electron Fermi energy changes with voltage the trion and neutral exciton shifts with voltage can be different~\cite{Mak2012}.

\subsection{Methods}
 
The excitons were generated by a semiconductor laser with excitation energy $E_{\rm ex}=1.96$~eV. Luminescence spectra were measured using a spectrometer with resolution 0.2~meV and a liquid-nitrogen-cooled CCD. The laser was either defocused or focused to a spot size $\sim 3.5$~$\mu$m. The IX lifetime was measured using the laser pulses of a rectangular shape with the duration 20~ns, period 80~ns, and edge sharpness $\sim 0.5$~ns, the emitted light was filtered by an interference filter and detected by a liquid-nitrogen-cooled CCD coupled to a PicoStar HR TauTec time-gated intensifier. The IX lifetimes measured at 2 K reach $\sim 10$~ns, orders of magnitude longer than the direct exciton lifetime in single-layer TMD~\cite{Korn2011}. The experiments were performed in a variable-temperature 4He cryostat at $T=1.7-293$~K.

The trion luminescence can be co-polarized with the incident light, counterpolarized, and unpolarized, as described in topical review~\cite{Bar-Joseph2005}. Since the polarization of the trion luminescence is a complex issue with multiple possible outcomes, it is hard to use it for a definitive assignment of a luminescence line to the trion. On the contrary, the temperature dependence distinguishes the interpretations of two lines in terms of trion and exciton vs. exciton and exciton, as outlined above. This is why the temperature dependence is used for the assignment of the luminescence line to the indirect trion in this work, polarization-resolved measurements form the subject for future studies. 

The studied van der Waals heterostructures were assembled using the standard peel-and-lift technique~\cite{Withers2015}. In brief, individual crystals of graphene, hBN, MoSe$_2$, and WSe$_2$ were first micromechanically exfoliated onto different Si substrates that were coated with a double polymer layer consisting of polymethyl glutarimide (PMGI) and polymethyl methacrylate (PMMA). The bottom PMGI was then dissolved releasing the top PMMA membrane with a target 2D crystal. Separately, a large graphite crystal was exfoliated onto an oxidized Si wafer, which later served as the bottom electrode. The PMMA membrane was then flipped over and aligned above an atomically-flat region of the graphite crystal using a micromechanical transfer stage. The two crystals were brought into contact and the temperature of the stage was ramped to $80^{\rm o}$~C in order to increase adhesion between the 2D crystals. Then, the PMMA membrane was slowly peeled off leaving the bilayer stack on the wafer. The procedure was repeated leading to a multi-crystal stack with the desired layer sequence. The long edges of MoSe$_2$ and WSe$_2$ monolayers were aligned (Fig.~S1). We note that the dependence on the angle beween the layers form the subject for future studies and do not affect the conclusions on neutral and charged IXs in this work. Finally, electron-beam lithography was employed to define contact regions to graphene and graphite crystals, which followed by metal evaporation of electrical contacts (3~nm Cr / 80~nm Au) and liftoff.

\end{suppinfo}

\bibliography{TMD_Ex_Tri_Full}

\end{document}